\documentclass[3p, 12pt, onecolumn]{elsarticle}
\usepackage{graphicx,amscd,amsmath,amssymb,verbatim}
\usepackage{amsfonts,epsfig}
\usepackage{mathptmx}
\usepackage{setspace}
\usepackage{epsfig}
\usepackage{url}
\usepackage{multirow}
\usepackage{subfigure}
\usepackage{graphicx,color}
\usepackage{algorithm,caption}
\usepackage{algpseudocode}
\usepackage{float}
\usepackage[normalem]{ulem}

\pdfoutput=1

\begin{document}

\journal{Physica A}

\begin{frontmatter}

\title{Crossover effects on the phase transitions phenomena translated by arborecences and spectral properties}

\author{Roberto da Silva}

\address{Instituto de F\'{i}sica, Universidade Federal do Rio Grande do Sul,
Porto Alegre Rio Grande do Sul, Brazil\\}


\begin{abstract}
This study investigates how visibility graphs constructed from Monte
Carlo Markov Chain time series of spin models capture the
critical behavior of the system. More precisely, we show that this
approach identifies continuous phase transitions as well as important
nuances, such as crossover effects occurring in the transition from a
critical line to a first-order line through a tricritical point, as
observed, for example, in the Blume--Emery--Griffiths model or, in a
simpler setting, in the Blume--Capel model. By applying Kirchhoff's
theorem, we show that the number of spanning trees of the resulting
graphs serves as a sensitive indicator of these phase transitions.
Furthermore, a qualitative analysis of the adjacency matrices based on
random matrix theory provides additional evidence for these phenomena.
The methodology developed here can potentially be extended to the
analysis of criticality in empirical time series from complex systems,
such as climate, financial, and epidemiological data, where the
Hamiltonian governing the dynamics is not necessarily known.

\end{abstract}

\end{frontmatter}

\section{Introduction}

\label{Section:Introduction}

One important extension of the Ising model to spin $S=1$ variables is the
well--known Blume--Emery--Griffiths (BEG) model \cite%
{BlumeEmeryGriffiths1971}, defined by the Hamiltonian 
\begin{equation}
\mathcal{H}=-J\sum_{\langle ij\rangle }s_{i}s_{j}-K\sum_{\langle ij\rangle
}s_{i}^{2}s_{j}^{2}+D\sum_{i}s_{i}^{2},\qquad s_{i}=0,\pm 1,
\end{equation}%
or, in its simpler form, the Blume--Capel (BC) model \cite{Blume,Capel},
which is obtained from the BEG Hamiltonian by setting $K=0$, 
\begin{equation}
\mathcal{H}_{\text{BC}}=-J\sum_{{\langle ij\rangle }}s_{i}s_{j}+D%
\sum_{i}s_{i}^{2}.
\end{equation}%
Here, $J$ denotes the bilinear exchange coupling, $K$ the biquadratic
interaction, and $D$ the crystal--field term, which controls the population
of non--magnetic states $s_{i}=0$.

Originally introduced to describe phase separation and superfluid ordering
in ${}^{3}$He--${}^{4}$He mixtures, the BEG model has since become a
paradigmatic framework for the investigation of tricritical phenomena. In
two dimensions, it exhibits a rich phase diagram comprising continuous phase
transitions in the Ising universality class, first--order transition lines,
and a tricritical point separating these two regimes \cite%
{MukamelBlume1974,BerkerWortis1976}. The coexistence of multiple relevant
couplings enhances crossover effects, which become particularly pronounced
in both analytical approaches and numerical simulations.

For moderate values of $D/J$ and $K/J$, the transition between the
paramagnetic and ferromagnetic phases is continuous and governed by the
two--dimensional Ising fixed point, characterized by the critical exponents $%
\nu =1$, $\beta =\tfrac{1}{8}$, and $\alpha =0$. As the parameters are
varied, this critical line terminates at a tricritical point, beyond which
the transition becomes first order. At tricriticality, the critical behavior
is described by a distinct set of exponents, $\nu _{t}=\tfrac{5}{9}$, $\beta
_{t}=\tfrac{1}{24}$, and $\alpha _{t}=\tfrac{8}{9}$, reflecting the presence
of an additional relevant scaling field \cite{MukamelBlume1974}.

The crystal--field parameter $D$ plays a central role in the crossover
properties of the BEG model. Large negative values of $D$ energetically
suppress the $s_{i}=0$ state, effectively reducing the system to the
standard spin--$\tfrac{1}{2}$ Ising model. In contrast, large positive
values of $D$ favor non--magnetic sites, leading to a strongly diluted
magnetic system. In two dimensions, this continuous variation in the density
of magnetically active spins induces a crossover between an Ising--like
regime and a lattice--gas--like diluted regime, thereby modifying the
structure of thermodynamic singularities such as the specific heat and
susceptibility. This density--driven crossover significantly complicates the
numerical identification of the true asymptotic critical behavior.

The proximity of the tricritical point leads to strong crossover effects. In
renormalization--group language, the parameter space of the BEG model
contains both an Ising fixed point and a tricritical fixed point, the latter
possessing more than one relevant direction. As a consequence, the effective
critical behavior observed depends on the length scale under consideration.

A characteristic crossover length 
\begin{equation}
\xi _{\times }\sim |g|^{-1/\phi }
\end{equation}%
separates the Ising--dominated regime from the tricritical one, where $g$
measures the distance from the tricritical point and $\phi $ is the
crossover exponent \cite{Lawrie}. For correlation lengths $\xi \ll \xi _{\times }$, the
system exhibits effective Ising--like critical behavior, whereas for $\xi
\gg \xi _{\times }$ the tricritical scaling regime becomes dominant. In
finite systems, or in simulations performed at parameters not asymptotically
close to criticality, this mechanism naturally leads to effective critical
exponents that interpolate between the Ising and tricritical values. Such
crossover effects can also be systematically investigated within the simpler
BC model.

Such crossover effects can alternatively be investigated using
time-dependent Monte Carlo (MC) simulations, where critical parameters and
exponents are extracted from the early stages of time evolution for both
short-range and long-range (mean-field) interactions in two and three
dimensions \cite{RdaSilvaPRE2002,RdaSilvaBJP2022,RdaSilvaPRE2022}. By
focusing on short-time dynamics, universal critical behavior can be accessed
without requiring full equilibration. Incorporating concepts from random
matrix theory further allows for a spectral characterization of criticality.
We have recently shown that matrices constructed from a small number of MC
steps, within the framework of Wishart ensembles, encode the critical
behavior of spin systems in their spectral properties \cite%
{RdasilvaIJMPC2023,RdasilvaIJMPC2024}. The same approach has been applied
along the critical line of the BC model; however, its ability to accurately
determine critical points becomes significantly affected near the
tricritical point, reflecting the influence of crossover phenomena \cite%
{Eliseu2024}.

The idea of constructing matrices from time series of physical observables,
such as the magnetization, whose spectra reflect the critical properties of
the underlying system, naturally opens the door to the use of further
concepts from the literature. One particularly appealing framework is the
visibility graph (VG), which maps time series into graphs \cite{Lacasa2008},
thereby enabling the application of complex network theory to the
investigation of temporal patterns with conceptual simplicity and strong
physical interpretability. In the visibility algorithm introduced by Lacasa 
\emph{et al.} \cite{Lacasa2008}, each data point $(t_{i},x_{t_{i}})$ is
represented as a node $i$, and two nodes $i$ and $j$ are connected if a
straight line drawn between $(t_{i},x_{t_{i}})$ and $(t_{j},x_{t_{j}})$ does
not intersect any intermediate data point. This construction allows one to
quantify temporal complexity topologically and to extract scaling properties
directly from the network structure.

Time series generated by spin models have already been explored within this
framework. For instance, Zhao \emph{et al.} \cite{Zhao2017} investigated
networks constructed from Ising model time series, while Gomez-Hernandez 
\emph{et al.} \cite{Gomez-Hernandez} demonstrated that networks built from
time series of both Ising and Kuramoto models retain the ability to detect
phase transitions. Similar analyses were reported by Moraes and Ferreira 
\cite{Ferreira} for the contact process. However, these studies did not
explore the spectral properties of random matrices associated with the
adjacency matrices, nor did they address a central question of the present
work: what can the number of spanning trees of graphs (or equivalently its
logarithm, the so-called structural entropy), constructed from distinct
MC evolutions of spin systems governed by the standard Metropolis
MCMC dynamics, reveal about phase transitions?

This idea was recently proposed by us in the context of the Ising model
\cite{SilvaArxiv}. In the present work, we extend and refine this method
to investigate the critical line of the BC model and its
associated crossover phenomena. In particular, we analyze both the
structural entropy—defined as the logarithm of any cofactor of the
Laplacian matrix, since all cofactors are identical according to
Kirchhoff's theorem—and the eigenvalue density and level-spacing
distribution of the adjacency matrix spectrum.

The remainder of this paper is organized as follows. In the next section,
we describe the methodology. Section \ref{Sec:Results} presents the main
results, and Sec. \ref{Sec:Conclusions} summarizes the conclusions.

\section{Methodology}

\label{Sec:Model}

We evolute the BC model using the heat-bath dynamics starting from a random
initial magnetization ($m_{0}\approx 0$ in general, or other value when
mentioned). Let us define 
\begin{equation}
m_{i,j}=\frac{1}{L^{2}}\sum_{k=1}^{L^{2}}s_{k}^{(i,j)}
\end{equation}%
as the magnetization per spin at the $i$-th MC step ($%
i=0,1,2,\ldots ,N_{\mathrm{steps}}-1$) of the $j$-th independent run ($%
j=1,\ldots ,N_{\mathrm{run}}$). Here, the spin variables $%
s_{k}^{(i,j)}=0,\pm 1$. Thus, we can construct the VG for the $%
j$-th evolution, represented by the time series $m_{0j},m_{1j},\ldots ,m_{N_{%
\mathrm{steps}}-1,j}$. The adjacency matrix of the VG is
defined by representing each data point $(i,m_{i,j})$ in the time series as
a node $i$. Two nodes $i$ and $i^{\prime }$ are said to be adjacent (or
connected) if a straight line can be drawn between the points $(i,m_{i,j})$
and $(i^{\prime },m_{i^{\prime },j})$ without intersecting any intermediate
points. A simple ilustration is shown for the case of Fig. \ref%
{Fig:ilustration}.

\begin{figure}[tbp]
\begin{center}
\includegraphics[width=1.0\columnwidth]{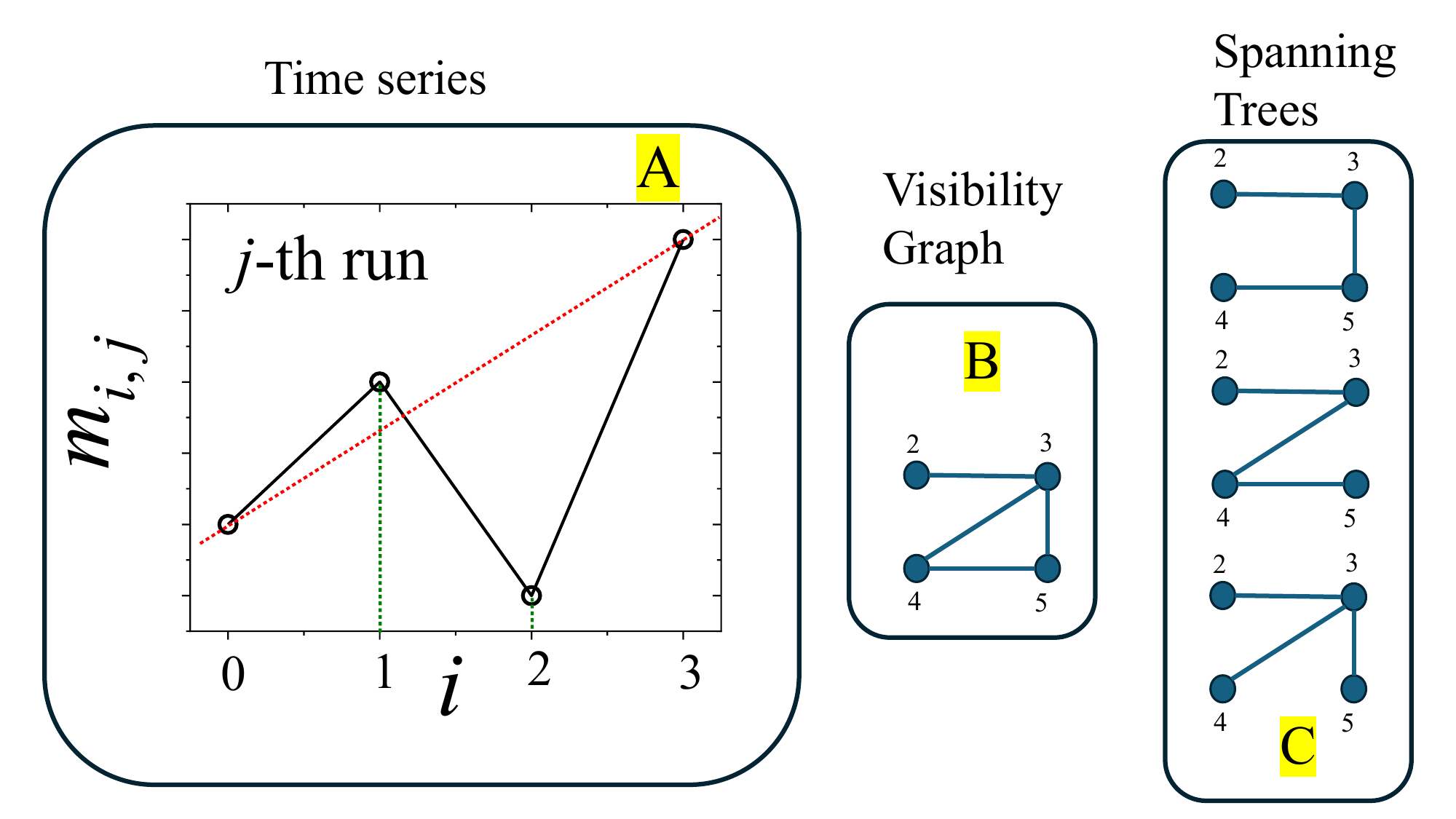}
\end{center}
\caption{\textbf{Plot A}: A possible time evolution ($j-$th evolution) of
the magnetization of the BC model for $N_{\text{steps}}=4$. \textbf{Plot B}:
Visibility graph corresponding to the time series. \textbf{Plot C}: All
spanning trees of the visiblity graph. }
\label{Fig:ilustration}
\end{figure}

Here we consider one arbitrary case with $N_{\mathrm{steps}}=4$. The Plot A
ilustrates the time evolution of magnetization for one run ($j$-th for the
example). Following the rule, node 0 is visible by 1. 1 is visible obvioulsy
by node 0 and by nodes 2 and 3. Finally, the node 3 is visible only by 2 and
1. The resulting vibility graph is shown in Plot B. One of most important
concepts of this paper is of spanning tree. Given a graph, a spanning tree
is a connected subgraph that includes all the nodes of the original graph
and contains no cycles. Trees, are in a general way a simpler form to cover
nodes of a way where all nodes are reachble for the other ones without
redundances. The number of spanning trees of a connected graph $G$, denoted
by $\tau (G)$, is a fundamental combinatorial quantity that appears in a
wide range of theoretical contexts (see for example \cite%
{Bollobas1998,Diestel2017,Stanley1999}).

A classical physical interpretation of $\tau (G)$ arises from its analogy
with electrical resistor networks, first established by Kirchhoff \cite{Kirchhoff1847}
in his pioneering work. Since simple graphs -- such as the VG
-- are fully characterized by their (symmetric) adjacency matrix $A$, whose
elements satisfy $a_{ij}=1$ if nodes $i$ and $j$ are connected and $a_{ij}=0$
otherwise, Kirchhoff formulated a fundamental result known as the \textit{%
Matrix--Tree Theorem} (see, for example, \cite{Harary1969}).

Given a connected graph $G$ (as is the case for VGs by
construction) with adjacency matrix $A$ , one can define another fundamental
matrix: the Laplacian matrix, given by 
\begin{equation*}
L=-A+D,
\end{equation*}%
where $D$ is the diagonal matrix whose elements are given by $%
D_{ii}=\sum_{j=1}^{N_{\mathrm{steps}}}a_{ij}$. By definition, $a_{ii}=0$.
All cofactors of the Laplacian matrix $L$ are equal, and each of them
corresponds to the number of spanning trees of the graph $G$. For
simplicity, we select the first cofactor, which yields 
\begin{equation}
\tau (G)=\det (L^{\prime })=\prod_{i=1}^{N_{\mathrm{steps}}-1}\lambda
_{i}(L^{\prime }),  \label{Eq:tauG_first}
\end{equation}%
where $L^{\prime }$ is the matrix obtained from $L$ by removing its first
row and first column. Its elements can be written as: 
\begin{equation*}
L_{ij}^{\prime }=%
\begin{cases}
\displaystyle\sum_{k=1}^{N_{\mathrm{steps}}}a_{i+1,k}, & \text{if }i=j, \\%
[1em] 
-\,a_{i+1,j+1}, & \text{otherwise,}%
\end{cases}%
\end{equation*}%
and $\lambda _{1},\lambda _{2},\ldots ,\lambda _{N_{\mathrm{steps}}-1}$
denote the eigenvalues of $L^{\prime }$.

Referring to Fig. \ref{Fig:ilustration}, let us clarify this point in a
pedagogical manner by explicitly examining the adjacency matrix of the graph
shown in Plot B and its corresponding Laplacian matrix: 
\begin{equation*}
A=\left( 
\begin{array}{cccc}
0 & 1 & 0 & 0 \\ 
1 & 0 & 1 & 1 \\ 
0 & 1 & 0 & 1 \\ 
0 & 1 & 1 & 0%
\end{array}%
\right) \text{, }L=\left( 
\begin{array}{cccc}
1 & -1 & 0 & 0 \\ 
-1 & 3 & -1 & -1 \\ 
0 & -1 & 2 & -1 \\ 
0 & -1 & -1 & 2%
\end{array}%
\right) \text{, and }L^{\prime }=\left( 
\begin{array}{ccc}
3 & -1 & -1 \\ 
-1 & 2 & -1 \\ 
-1 & -1 & 2%
\end{array}%
\right)
\end{equation*}

We observe that the determinant of $L^{\prime }$ is equal to 3, which is
naturally consistent with its eigenvalues, $\lambda _{1}=\sqrt{3}+2$, $%
\lambda _{2}=2-\sqrt{3}$, and $\lambda _{3}=3$, since $\lambda _{1}\lambda
_{2}\lambda _{3}=3$. Therefore, the graph has exactly three distinct
spanning trees, which are shown in Plot C of Fig. \ref{Fig:ilustration}.

An important observation is that, if we consider the eigenvalues of the
Laplacian matrix $L$ (rather than those of the reduced matrix $L^{\prime }$%
), denoted by $\xi _{1},\xi _{2},\ldots ,\xi _{N_{\mathrm{steps}}}$, then
necessarily $\xi _{1}=0$, while all the remaining eigenvalues are strictly
positive, provided the graph is connected. In this case, one can show that:

\begin{equation}
\tau (G)=\frac{1}{N_{steps}}\prod_{i=2}^{N_{\mathrm{steps}}}\xi _{i}(L)
\label{Eq:tauG_second}
\end{equation}%
and in this case you do not think in coffactors to calculate $\tau (G)$, but
it is only an alternative and both directions (Eqs. \ref{Eq:tauG_first} and %
\ref{Eq:tauG_second}) are valid as you want.

Within this approach, the computation of cofactors becomes unnecessary for
evaluating $\tau (G)$. This representation is merely an alternative
formulation, and both expressions (Eqs. \ref{Eq:tauG_first} and \ \ref%
{Eq:tauG_second}) are mathematically equivalent and can be employed
according to convenience.

Having illustrated the method with this simple example, we now turn to the
more general case that will be addressed throughout this manuscript: a
longer magnetization time series. In this scenario, the adjacency matrix $A$
of the associated time VG can be written, in general form, as

\begin{equation*}
a_{i,i^{\prime }}=\prod_{k=i+1}^{i^{\prime }-1}H\left( m_{i,j}+(m_{i^{\prime
},j}-m_{i,j})\cdot \frac{(k-i)}{(i^{\prime }-i)}-m_{k,j}\right)
\end{equation*}%
where $H(x)$ is the Heaviside step function:%
\begin{equation*}
H(x)=\left\{ 
\begin{array}{ccc}
0 & \text{if} & x<0 \\ 
&  &  \\ 
1 & \text{if } & x\geq 0.%
\end{array}%
\right.
\end{equation*}%
In other words, for node $i$ to be adjacent to node $i^{\prime }$, it is
necessary that $H=1$ for all $k$ in the interval $i+1\leq k$ $\leq i^{\prime
}-1$. The matrix $A$ is therefore a binary, symmetric matrix with a null
diagonal.

At first glance, one might be tempted to argue that VGs
constructed from magnetization time series at high temperatures should
resemble random graphs. In the high--temperature regime, the magnetization
behaves approximately as non-correlated Gaussian noise, and one could naively expect that
the associated adjacency matrices would share properties with symmetric
random matrices. In classical random graph models, two nodes $i$ and $j$ are
connected independently with probability $p$, and disconnected with
probability $1-p$. In suitable limits, the spectral properties of such
matrices may approach those described by the Gaussian Orthogonal Ensemble
(GOE), including the emergence of the Wigner semicircle law (or controlled
deviations from it) for the eigenvalue density.

However, this reasoning is incorrect in the present context. Even if the
magnetization time series becomes Gaussian at high temperatures, the
corresponding VG is not a random graph in the Erd\H{o}s--R\'{e}%
nyi sense. The crucial difference lies in the construction rule: edges in a
VG are not assigned independently with a fixed probability,
but are determined by deterministic geometric constraints imposed by the
time series. As a consequence, the entries of $A$ are strongly correlated,
and the resulting network ensemble is structurally distinct from classical
random graph ensembles.

This distinction is essential. The eigenvalue density and level spacing
distribution of adjacency matrices derived from VGs should not
be directly compared with GOE predictions at any temperature. Instead, an
appropriate reference must be built from visibility graphs generated from
Gaussian noise itself, preserving the same construction mechanism. In the
next section, we present our results.

\section{Results}

\label{Sec:Results}

Thus, we perform MC simulations of the two-dimensional BC model
using heat-bath dynamics. The system is evolved for $N_{\text{steps}}=100$
time steps, starting from initial conditions with $m_{0}\approx 0$ (except
in cases where other initial conditions are explicitly specified), for each
value of $D$. Each evolution is repeated over $N_{\text{run}}=10^{5}$
independent runs. 

For every run, the resulting time series generates a VG with
$N_{\text{steps}}=100$ nodes, together with its corresponding adjacency and
Laplacian matrices. This procedure yields $10^{7}$ eigenvalues for the
estimation of the eigenvalue density and $10^{5}$ different values of the
number of spanning trees, allowing a reliable average for estimating the
structural entropy per node, or simply the tree entropy, defined as
\cite{SilvaArxiv}
\begin{equation*}
s=\lim_{N_{\text{run}}\rightarrow \infty ,\, N_{\text{steps}}\rightarrow \infty}
\frac{1}{N_{\text{steps}}N_{\text{run}}}
\sum_{i=1}^{N_{\text{run}}}\ln \tau(G_{i}) .
\end{equation*}

We considered five different values of $D/J$ in our simulations: $0$, $1$,
and $1.5$, which are points far from the tricritical point, as well as
$D/J=1.9501$ and the tricritical point itself, $D/J=1.9655$. These values
are listed in Table~\ref{Table:critical_points} (see \cite{Butera,DeBeale}). 

For simplicity, in tables and plots we use $D$ instead of $D/J$, which is
equivalent to setting $J=1$ without loss of generality. 

\begin{table}[tbp] 
\centering
$
\begin{tabular}{llllll}
\hline\hline
$D_{C}$ & $0$ & $1$ & $1.5$ & $1.9501$ & $1.9655$ \\ \hline
$T_{C}$ & $1.695$ & $1.398$ & $1.150$ & $0.650$ & $0.610$ \\ \hline\hline
\end{tabular}
$
\caption{Critical points of BC model choosen to be studied
in this work. For reference see for example \cite{Butera,DeBeale}.}\label%
{Table:critical_points}
\end{table}

Thus, we divide our results into two parts. The first part concerns the
Laplacian matrix, where we compute the number of spanning trees of the
VGs constructed from the magnetization time evolutions of
the BC model. The second part consists of a random matrix
analysis of the adjacency matrix of the corresponding VGs.
Although this second approach does not provide a direct method to locate
the critical points, as the first one does, it offers useful supporting
insights, particularly through the study of crossover effects and through
the characterization of the low- and high-temperature regimes revealed by
the VGs.

\subsection{Laplacian matrix results}

\begin{figure}[tbp]
\begin{center}
\includegraphics[width=0.75\columnwidth]{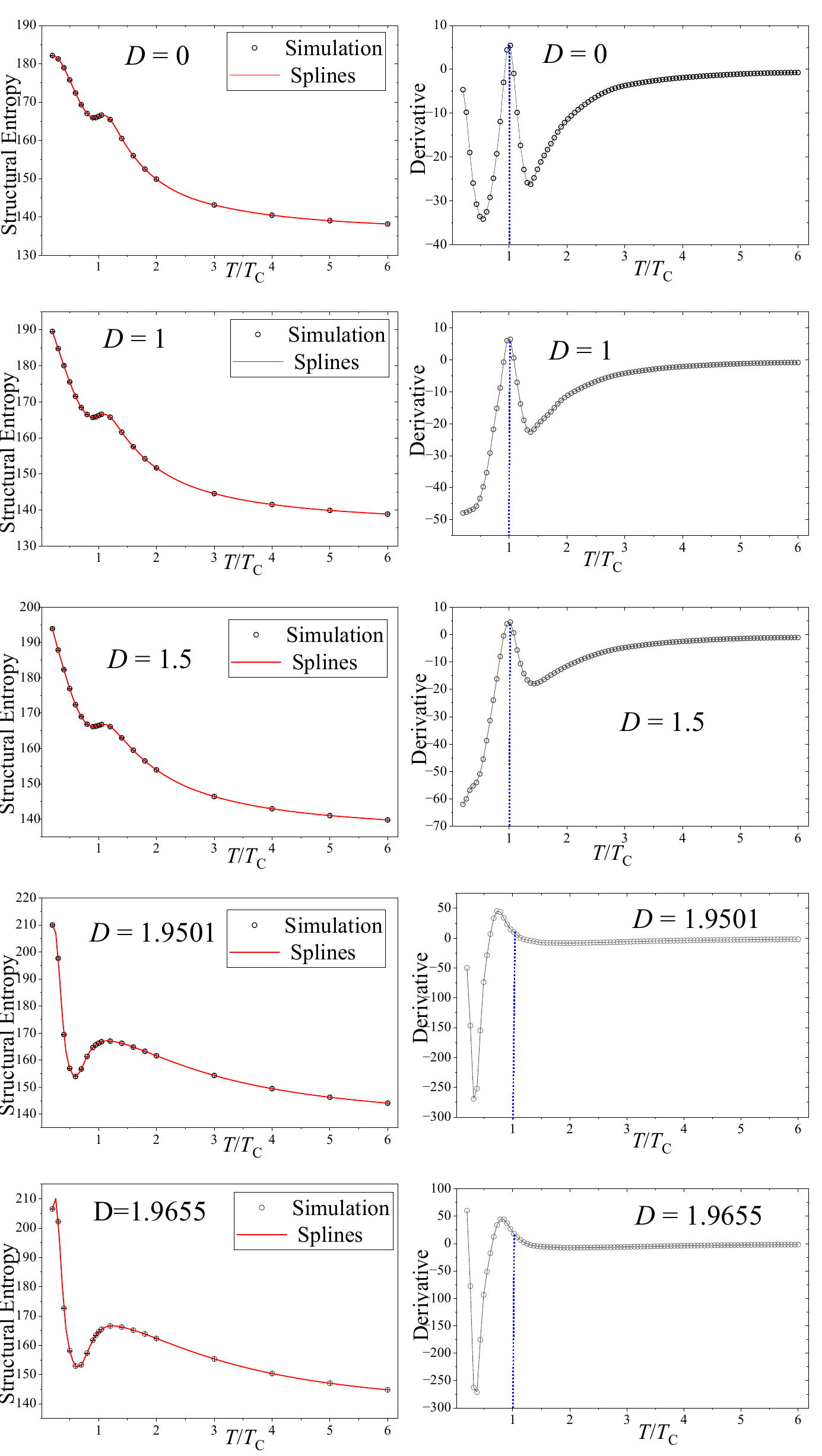}
\end{center}
\caption{Structural entropy and its derivative as a function of $T/T_{C}$ for
different points along the critical line of the BC model.
The tree entropy (structural entropy) exhibits a plateau in the
vicinity of the critical temperature. In contrast, the derivative
of this quantity shows a pronounced peak at $T=T_{C}$ for points
far from the tricritical point. In the vicinity of the tricritical
point ($D_{C}=1.9501$) and at the tricritical point itself
($D_{C}=1.9655$), crossover effects affect the estimates, as expected. }
\label{Fig:Structural_Entropy_and_derivative}
\end{figure}

We begin by presenting the structural entropy, estimated as
$S = N_{\text{steps}} s = \frac{1}{N_{\text{run}}}
\sum_{i=1}^{N_{\text{run}}}\ln \tau(G_{i})$, as a function of $T/T_{C}$.
In Fig.~\ref{Fig:Structural_Entropy_and_derivative} we show the structural
entropy (left panels) together with its derivative (right panels) for the
different values of $D$ (see Table~\ref{Table:critical_points}). The red
curve corresponds to a spline interpolation included only as a guide to
the eye.

In the left panels, a small plateau around $T/T_{C}=1$ can be observed for
the first three values ($D=0$, $1$, and $1.5$), since these points are far
from the tricritical region and therefore do not exhibit significant
crossover effects. In the corresponding right panels, the derivative of this quantity
consistently exhibits a peak at the critical temperature, as also
observed for the spin-$1/2$ Ising model \cite{SilvaArxiv}.
However, these peaks do not coincide with $T/T_{C}=1$ for $D=1.9501$, which
is close to the tricritical point, nor for the tricritical point itself
($D=1.9655$), illustrating that crossover effects affect the estimates, as
expected.

Following our analysis, it is important to emphasize that the tree entropy
does not exhibit a maximum at the critical temperature; instead, the
maximum appears in its derivative. However, as observed for the spin-$1/2$
Ising model \cite{SilvaArxiv}, a maximum occurs near $T=T_{C}$ when
$m_{0}\neq 0$. When $m_{0}$ becomes sufficiently small, this peak
gradually transforms into a plateau that contains $T=T_{C}$. The same
behavior is observed here for the critical points of the BC model
with different anisotropies.

\begin{figure}[tbp]
\begin{center}
\includegraphics[width=0.8\columnwidth]{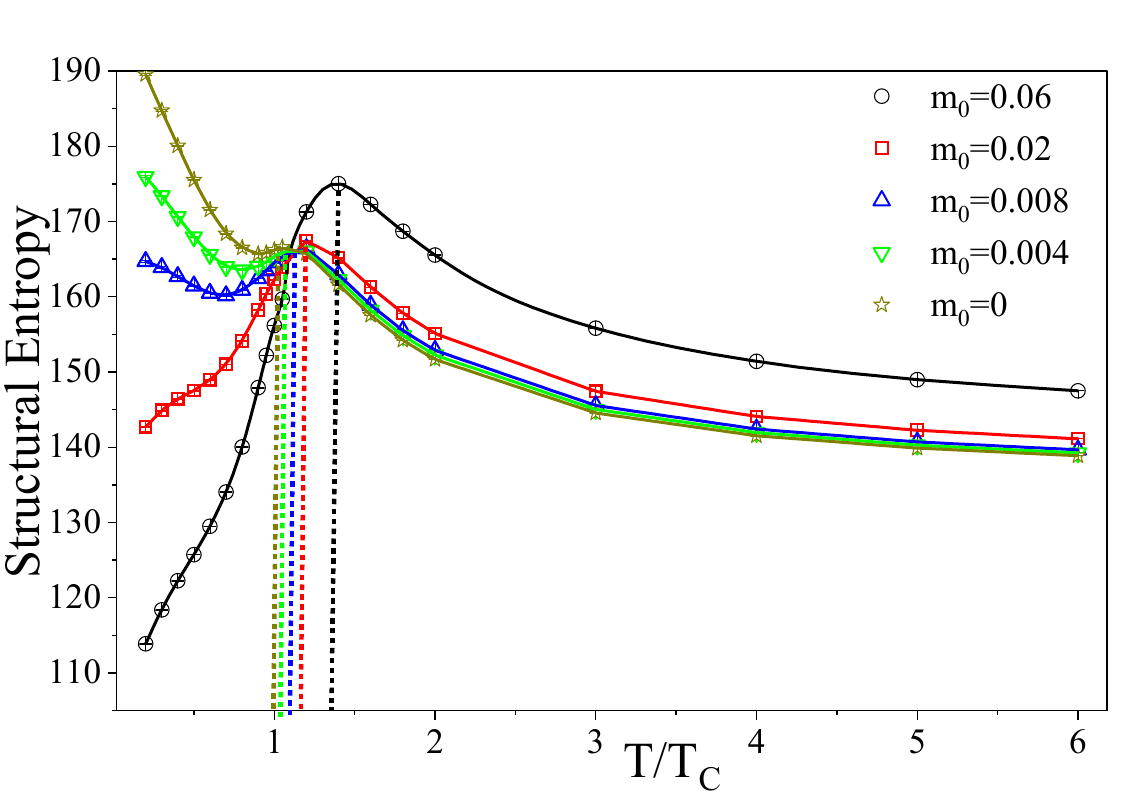}
\end{center}
\caption{Effects of the initial conditions on the tree entropy. As $m_{0}\to 0$,
the peak gradually transforms into a plateau that includes the value
$T=T_{C}$.}
\label{Fig:Effect_of_inicial_condition}
\end{figure}

To illustrate this effect, we show the evolution of the peak as a function
of $m_{0}$. For simplicity, we present the case $D=1$, since the other
cases behave similarly, displaying the structural entropy for
$m_{0}=0.06$, $0.02$, $0.008$, and $0.004$ (see
Fig.~\ref{Fig:Effect_of_inicial_condition}). One can observe that the
entropy exhibits a maximum at $\tau^{\ast}=T^{\ast}/T_{C}$; however, this
maximum does not coincide with $\tau=1$ when $m_{0}\neq 0$. On the other hand,
As $m_{0}\to 0$, the peak gradually transforms into a plateau that includes 
the value $T=T_{C}$, by showing that the phenomenon should correspond to a maximum-entropy
behavior, although with a more complex explanation. For clarity, we 
use $\tau$ and $T/T_{C}$ interchangeably in this work, since both denote the 
same quantity.

\begin{figure}[tbp]
\begin{center}
\includegraphics[width=0.9\columnwidth]{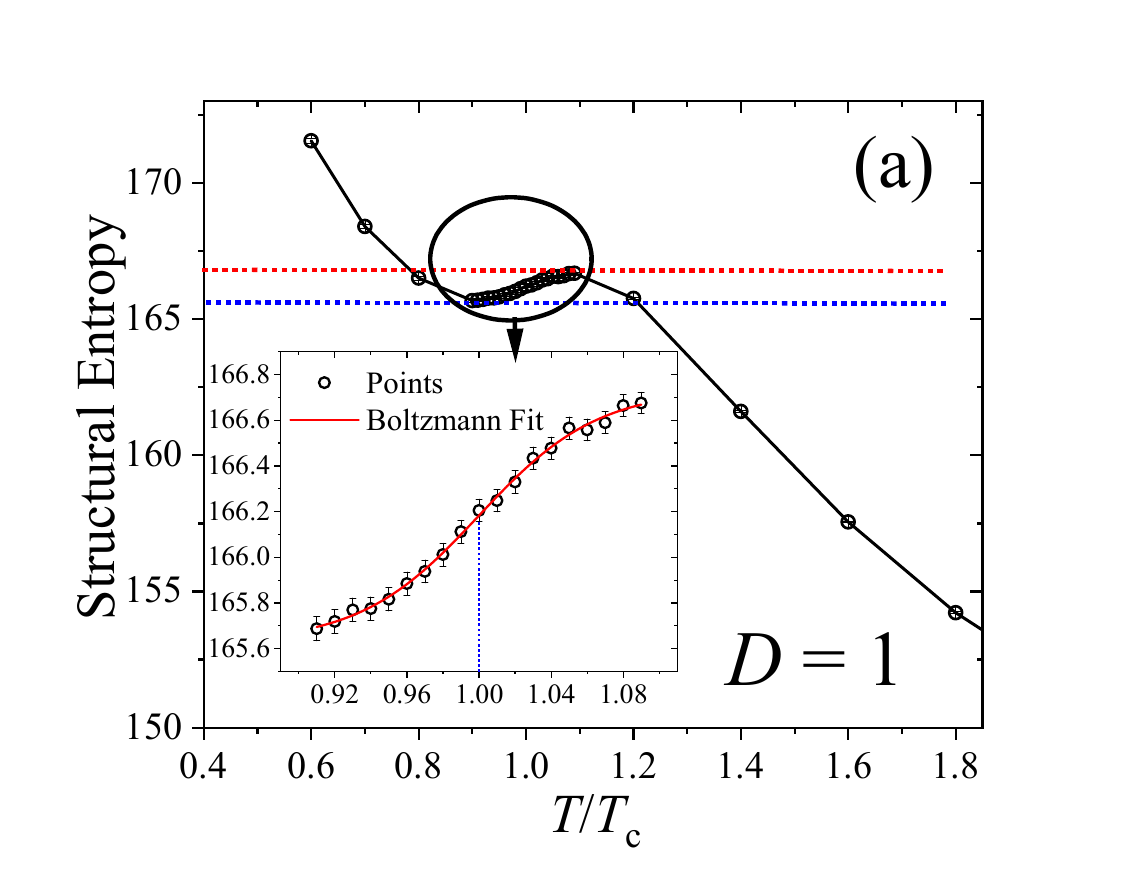} %
\includegraphics[width=0.9\columnwidth]{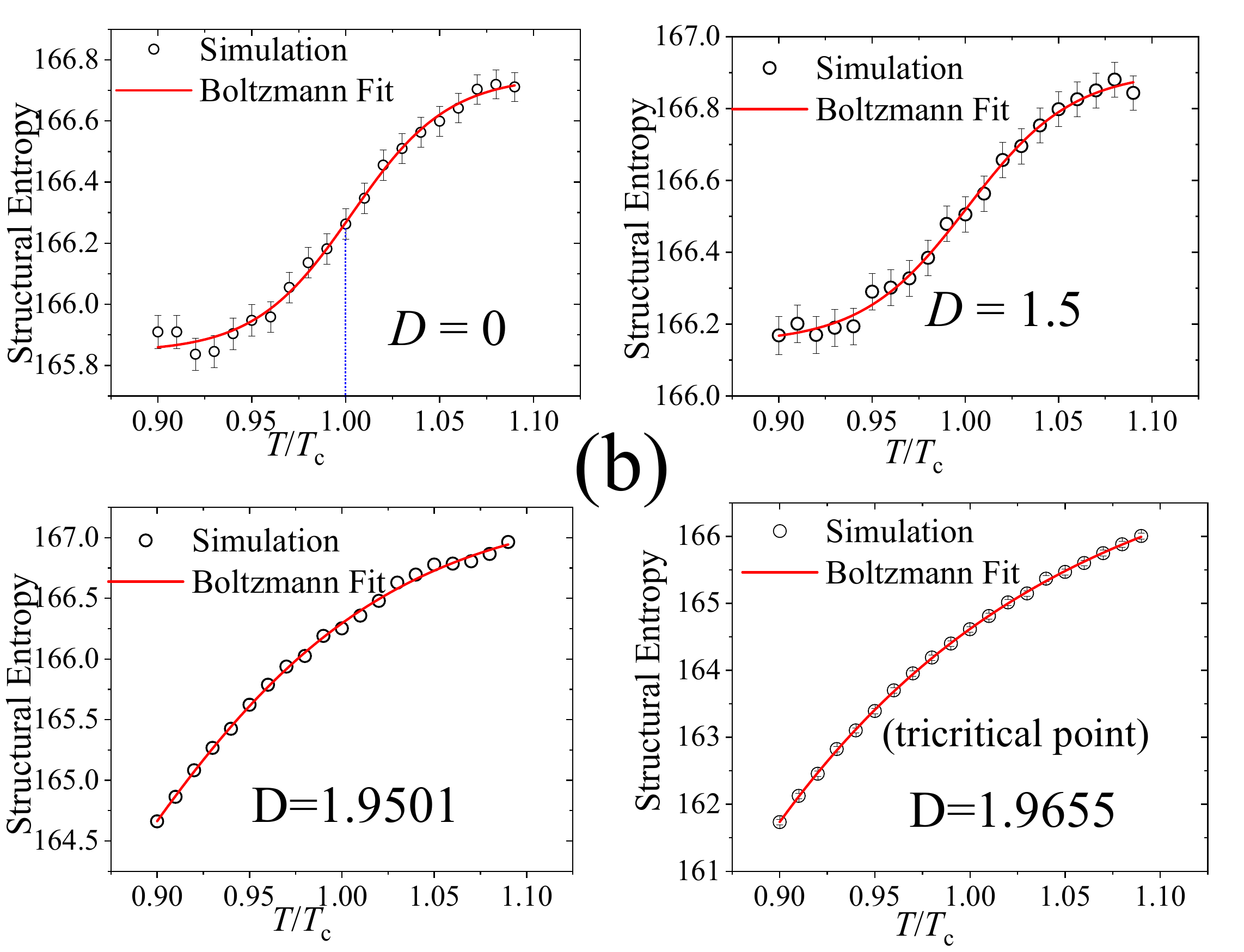}
\end{center}
\caption{(a) Zoom of the plateau region for the case $D=1$. This region is well
described by a Boltzmann (sigmoidal) function. (b) Boltzmann fits for the
other values of $D$. The tricritical point and the nearby point are not
well fitted due to crossover effects, which are characterized by the
parameter $x_{0}$.%
}
\label{Fig:refinement}
\end{figure}

It is important to refine the plateau region. By zooming into this part of
the figure, we obtain the behavior shown in Fig.~\ref{Fig:refinement}(a),
which reveals a sigmoidal dependence of the entropy as a function of
$\tau$. This suggests that it is natural to perform a fit using a
Boltzmann function:
\begin{equation*}
S=\frac{A_{1}-A_{2}}{1+e^{\frac{(\tau -\tau _{C})}{\sigma _{\tau }}}}+A_{2},
\end{equation*}
as illustrated in the inset of Fig.~\ref{Fig:refinement}(a).

It is useful to clarify the meaning of the fitting parameters. The
constants $A_{1}$ and $A_{2}$ represent the asymptotic values of $S$ in the
two limiting regimes: $A_{1}$ corresponds to the asymptotic value for
$\tau \ll \tau_{C}$ (initial plateau), whereas $A_{2}$ corresponds to the
asymptotic value for $\tau \gg \tau_{C}$ (final plateau). The parameter
$\sigma_{\tau}$ represents the transition width, characterizing how
abruptly the system changes between the two regimes. Finally,
$\tau_{C}$ indicates the location of the transition between these regimes.

It is important to note that for $\tau=\tau_{C}$ one obtains
$S(\tau=\tau_{C})=\frac{A_{1}+A_{2}}{2}$, and the maximal slope occurs
exactly at
\begin{equation*}
\left. \frac{dS}{d\tau }\right|_{\tau_{C}}=\frac{A_{2}-A_{1}}{4\sigma_{\tau}} .
\end{equation*}
Moreover, by computing
\begin{equation*}
\frac{d^{2}S}{d\tau^{2}}=
\frac{(A_{2}-A_{1})\left(e^{\frac{\tau-\tau_{C}}{\sigma_{\tau}}}
- e^{2\frac{\tau-\tau_{C}}{\sigma_{\tau}}}\right)}
{\sigma_{\tau}^{2}\left(1+e^{\frac{\tau-\tau_{C}}{\sigma_{\tau}}}\right)^{3}},
\end{equation*}
we observe that for $\tau=\tau_{C}$ one has $\frac{d^{2}S}{d\tau^{2}}=0$,
showing that this point is indeed an inflection point.

The fit yields, with a coefficient of determination $\mathrm{COD}=0.9976$,
the values $A_{1}=165.6\pm0.03$, $A_{2}=166.7\pm0.03$,
$\sigma_{\tau}=0.033\pm0.002$, and $\tau_{C}=0.9998\pm0.0019$, which
indeed includes $\tau_{C}=\frac{T}{T_{C}}=1$ within the confidence
interval. 

We repeat the same procedure for other values of $D$ in the refinement
region, as shown in Fig.~\ref{Fig:refinement}(b). A good Boltzmann fit is
obtained for $D=0$ and $D=1.5$, exactly as observed for $D=1$, with the
confidence interval again containing $\tau_{C}=\frac{T}{T_{C}}=1$. This is
not the case for the point near the tricritical value nor for the
tricritical point itself. The fitting results are summarized in
Table~\ref{Table:refinement}. Notably, we obtain $\tau_{C}\approx1$ for
points far from the tricritical point. On the other hand, the results show
that the sigmoidal fit is not appropriate for points close to the
tricritical region.

\begin{table}[tbp] \centering%
$%
\begin{tabular}{lll}
\hline\hline
$D_{C}$ & $\tau _{C}$ & COD \\ \hline
$0$ & $1.0031\pm 0.0022$ & $0.9944$ \\ 
$1$ & $0.9998\pm 0.0018$ & $0.9979$ \\ 
$1.5$ & $1.0012\pm 0.0023$ & $0.9951$ \\ 
$1.9501$ & $0.856\pm 0.051$ & $0.9984$ \\ 
$1.9655$ & $0.44\pm 0.56$ & $0.9998$ \\ \hline\hline
\end{tabular}%
$%
\caption{Values of $\tau_{C}$ and the coefficient of determination obtained from
the Boltzmann fit applied to the refinement region.}\label{Table:refinement}%
\end{table}%

\subsection{Random matrices theory}

We now complete our analysis by examining the results in the framework of
random matrix theory (RMT). In this part, we do not aim to locate the critical
temperature; instead, we investigate how temperature affects the density
of eigenvalues and the spacing distribution.

\begin{figure}[tbp]
\begin{center}
\includegraphics[width=1.0\columnwidth]{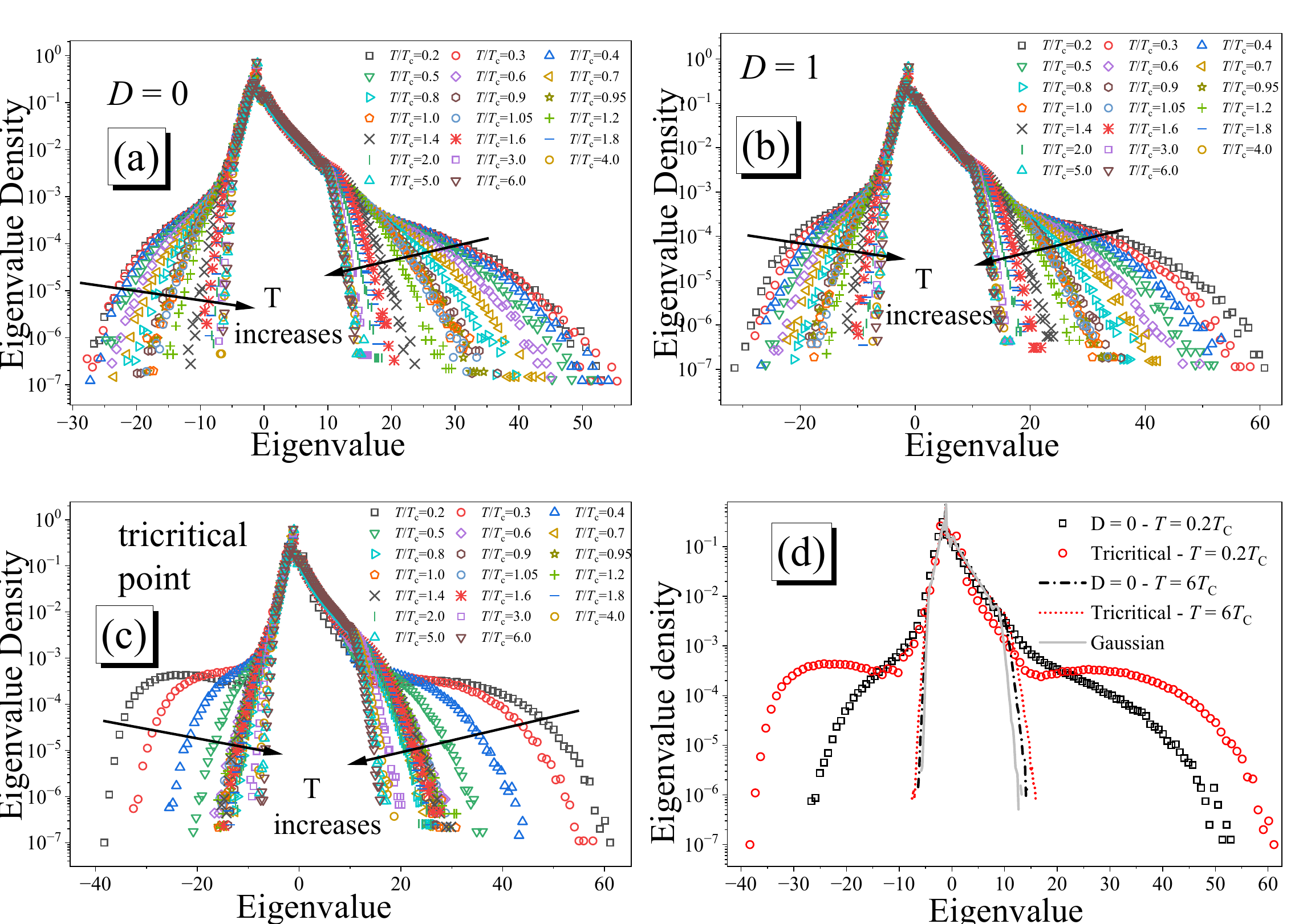}
\end{center}
\caption{Density of eigenvalues of the adjacency matrix of the VGs
for different temperatures. Panels (a) and (b) correspond to critical
points, while panel (c) corresponds to the tricritical point. Panel (d)
shows a comparison between the critical and tricritical cases at low and
high temperatures, together with the eigenvalue density of a visibility
graph obtained from uncorrelated Gaussian noise.}
\label{Fig:Density_of_eigenvalues}
\end{figure}

We begin by analyzing the eigenvalue density of the adjacency matrices of
the VGs constructed for different anisotropies $D$. For
simplicity, we focus on the cases $D=0$, $D=1$, and the tricritical point.
The results are shown in Fig.~\ref{Fig:Density_of_eigenvalues}(a), (b), and
(c), respectively.

We observe a tendency in which lower temperatures correspond to heavier
tails in the eigenvalue density. Although the density of states appears to
behave similarly at the critical and tricritical points, clear differences
emerge at low temperatures. This can be seen in
Fig.~\ref{Fig:Density_of_eigenvalues}(d), where we illustrate the case
$T=0.2T_{C}$ for both a critical point ($D=0$) and the tricritical point.

In the same panel, we also observe that at high temperatures both cases
tend to approach the density of states obtained from uncorrelated Gaussian
noise (represented by the gray line). The gray curve is clearly not the
semicircle law; rather, it corresponds to a stable law obtained when
deriving the density of states of adjacency matrices of VGs
constructed from uncorrelated Gaussian noise. To the best of our knowledge,
there are no results in the literature describing stable laws for the
eigenvalue density of VGs, suggesting that this point
deserves further investigation.

Finally, it is important to note that the spectrum at the critical
temperature exhibits, as expected, an intermediate tail between the
low-temperature (longer tails) and high-temperature (shorter tails)
regimes.

\begin{figure}[tbp]
\begin{center}
\includegraphics[width=1.0\columnwidth]{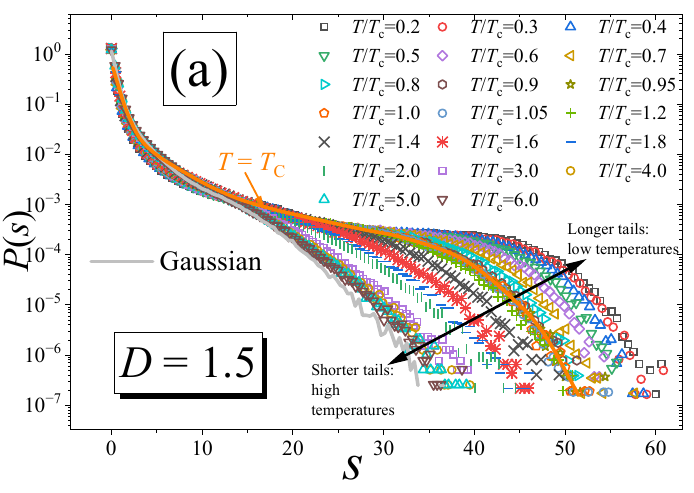} %
\includegraphics[width=1.0\columnwidth]{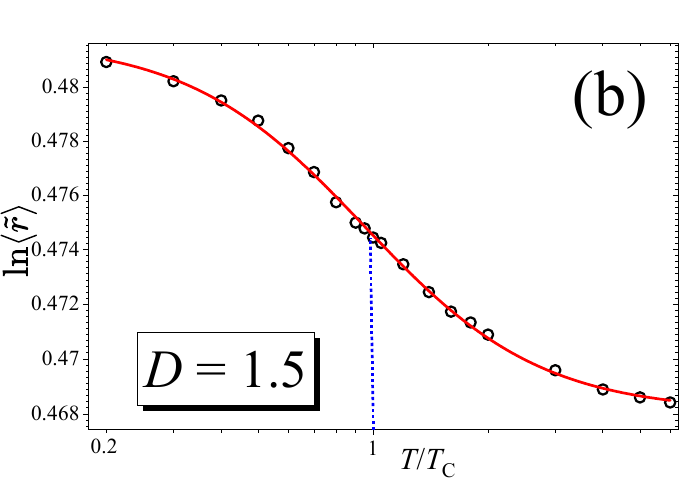} 
\end{center}
\caption{Spacing distribution of the eigenvalues of adjacency matrices of
VGs. (a) Spacing distribution for $D=1.5$. (b) Average value
of $\tilde{r}$ for $D=1.5$ as a function of $T/T_{C}$, which can be used
instead of the spacing distribution.}
\label{Fig:Spacing}
\end{figure}

Next, we focus on the spacing distribution in order to investigate the
effects of temperature for fixed anisotropies. We first order the
eigenvalues and then compute histograms of the spacings
$s=\frac{\lambda_{i}-\lambda_{i-1}}{D}$, with $i=2,\ldots,N_{\text{steps}}$.
We begin by considering an anisotropy far from the tricritical point
($D=1.5$), as shown in Fig.~\ref{Fig:Spacing}(a).

Here the same pattern observed for the eigenvalue density is also
established: longer tails correspond to low temperatures, while shorter
tails correspond to high temperatures, which tend to approach the spacing
distribution of the VG constructed from uncorrelated
Gaussian noise (gray line). The critical temperature corresponds to an
intermediate tail.

At this point it is useful to consider a simple and interesting ratio
proposed by Oganesyan and Huse \cite{Oganesyan}:
\begin{equation*}
\tilde{r}_{n}=\frac{\min (s_{n},s_{n-1})}{\max (s_{n},s_{n-1})}
=\min \left(r_{n},\frac{1}{r_{n}}\right)
\end{equation*}
where
\begin{equation*}
r_{n}=\frac{s_{n}}{s_{n-1}}
\end{equation*}
with $s_{n}=\lambda_{n+1}-\lambda_{n}$. This quantity does not require unfolding because ratios of consecutive
level spacings are independent of the local density of states.
Therefore, it allows a simpler comparison with experiments than the
traditional level-spacing distribution \cite{Atas}. We plot
$\tilde{r}$ as a function of $T/T_{C}$ for the case $D=1.5$ in
Fig.~\ref{Fig:Spacing}(b). The results are presented on a semi-log scale.

We observe that the values of $\langle \tilde{r} \rangle$ range from
approximately $0.468$ up to slightly above $0.48$. It is interesting to
compare these values with those obtained for standard random-matrix
ensembles. For example, for the Poisson ensemble the distribution of
$\tilde{r}$ is given by
\begin{equation*}
P(\tilde{r})=\frac{2}{(1+\tilde{r})^{2}}
\end{equation*}
for $0\leq \tilde{r}\leq 1$, and zero otherwise, which leads to
\begin{equation*}
\langle \tilde{r}\rangle_{\text{Poisson}}
=2\int_{0}^{1}\frac{\tilde{r}}{(1+\tilde{r})^{2}}\,d\tilde{r}
=2\ln 2-1\approx 0.3863 .
\end{equation*}

On the other hand, the random-matrix ensembles GOE, GUE, and GSE
(Gaussian orthogonal, unitary, and symplectic ensembles), corresponding
to $\beta=1,2,$ and $4$, lead to the distribution (see
Atas \textit{et al.} \cite{Atas})
\begin{equation*}
P(\tilde{r})=
C_{\beta}\frac{(\tilde{r}+\tilde{r}^{2})^{\beta}}
{(1+\tilde{r}+\tilde{r}^{2})^{1+\frac{3}{2}\beta}}
\end{equation*}
for $0\leq \tilde{r}\leq 1$, and zero otherwise, where $C_{\beta}$ is the
normalization constant. The corresponding mean values are
\begin{equation*}
\langle \tilde{r}\rangle_{\beta=1}
=4-2\sqrt{3}\approx 0.5359,
\end{equation*}
\begin{equation*}
\langle \tilde{r}\rangle_{\beta=2}
=\frac{2\sqrt{3}}{\pi}-\frac{1}{2}\approx 0.6027,
\end{equation*}
and
\begin{equation*}
\langle \tilde{r}\rangle_{\beta=4}
=\frac{32}{15}\frac{\sqrt{3}}{\pi}-\frac{1}{2}\approx 0.6762 .
\end{equation*}

Therefore, the values obtained at the critical point lie between those
predicted for the Poisson and GOE ensembles. The same behavior is observed
for the other critical points studied in this work, namely $D=0$ and
$D=1$.

\begin{figure}[tbp]
\begin{center}
\includegraphics[width=1.0\columnwidth]{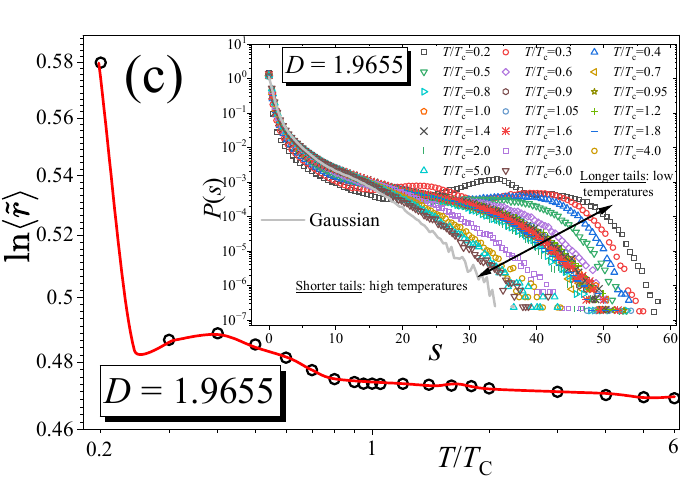}
\end{center}
\caption{$\langle \tilde{r} \rangle$ for the tricritical point as a function of
$T/T_{C}$. The inset shows the corresponding spacing distribution for
this point.}
\label{Fig:spacing_tricritical}
\end{figure}

On the other hand, for the tricritical point at low temperatures we observe
a slight increase in $\langle \tilde{r} \rangle$, reaching
$\langle \tilde{r} \rangle \approx 0.58$, which is larger than
$\langle \tilde{r} \rangle_{\beta=1}$ but still smaller than
$\langle \tilde{r} \rangle_{\beta=2}$, as shown in
Fig.~\ref{Fig:spacing_tricritical}. 

For $T\rightarrow\infty$, we obtain
$\langle \tilde{r} \rangle_{\infty} \approx 0.468$ for both the critical
and tricritical cases. This value corresponds to the spacing distribution
of the spectrum of a VG constructed from uncorrelated
Gaussian noise. The inset shows the spacing distribution for the
tricritical case, highlighting the anomalous behavior observed at low
temperatures.

\section{Summaries, conclusions, and discussions}

\label{Sec:Conclusions}

We present a study of the universality of the Blume--Capel (BC) model by using
spectral properties of visibility graphs (VGs)constructed from time series
generated by Monte Carlo simulations. First, we calculate the eigenvalues
of the Laplacian to show that the logarithm of the number of spanning trees
of the visibility graphs (tree entropy, or structural entropy as we called
it in \cite{SilvaArxiv}) is a good identifier of critical phenomena and of
the crossover effects present in this model.

For this purpose, it is necessary to calculate the derivative of this
quantity, which presents a pronounced peak at the critical temperature for
points far from the tricritical point. Alternatively, a refinement of the
region of structural entropy where the critical temperature is located
indicates a good fit with a Boltzmann function, showing that the inflection
point occurs exactly at $T=T_{C}$.

In a second analysis, we study the eigenvalue density and the spacing
distribution of the adjacency matrices of the VGs generated
by the Monte Carlo evolutions. Such an analysis illustrates that at high
temperatures both the density and the spacing distributions exhibit
shorter tails, whose limit ($T\rightarrow\infty$) corresponds to the
spectrum of a visibility graph generated from uncorrelated noise. At low
temperatures, correlations lead to longer tails.

By using the interesting ratio $\tilde{r}$ proposed by Oganesyan and Huse,
which condenses the information of the spacing distribution, we show that
the average value $\langle \tilde{r}\rangle$ lies between the values
predicted for the Poisson and the Gaussian orthogonal ensemble (GOE).
However, the crossover effects drastically modify the behavior at low
temperatures, where the average ratio surpasses the GOE value. At high
temperatures, this statistic corroborates the limit
$\langle \tilde{r}\rangle_{\infty}\approx 0.468$, corresponding to the
spectrum of a VG constructed from uncorrelated Gaussian
noise.

Although initially applied to the Ising model \cite{SilvaArxiv}, and here
extended with several additional results to cover the nuances of the
BC model (an Ising model with spin $1$ and anisotropy), the
method can be applied to time series from a variety of problems to
localize critical behavior in phenomena where the Hamiltonian is not
known, such as climate, financial, epidemic time series, chaotic systems,
and others. We believe that this powerful method can point to important
directions and lead to new discoveries in complex systems.

\textbf{Acknowledgements}

R. 	da Silva acknowledges the financial support of CNPq under grants 304575/2022‑4 and 406820/2025‑2.
This research was carried out with the support of the high‑performance computing resources provided by the Information Technology Superintendence of the University of São Paulo and by the Lovelace Cluster at IF‑UFRGS, Brazil.


\begin{thebibliography}{99}
\bibitem{BlumeEmeryGriffiths1971} M. Blume, V. J. Emery, R. B. Griffiths,
Ising Model for the $\lambda $ Transition and Phase Separation in $^{3}$He-$%
^{4}$He Mixtures, Phys. Rev. A \textbf{4}, 1071--1077 (1971)

\bibitem{Blume} M. Blume, Theory of the First-Order Magnetic Phase Change in
UO2, Phys. Rev. \textbf{141}, 517--524 (1966)

\bibitem{Capel} H. W. Capel, On the possibility of first-order phase
transitions in Ising systems of triplet ions with zero-field splitting I and
II, Physica \textbf{32}, 966--988 (1966), \textbf{33}, 295--314 (1967)

\bibitem{MukamelBlume1974} D. Mukamel, and M. Blume, Critical behavior of
the Blume--Emery--Griffiths model, Phys. Rev. A \textbf{10}, 610--616 (1974)

\bibitem{BerkerWortis1976} A. N. Berker, M. Wortis,
Blume--Emery--Griffiths--Potts model in two dimensions, Phys. Rev. B \textbf{%
14}, 4946--4963 (1976)

\bibitem{Lawrie} I. D. Lawrie and S. Sarbach, Theory of Tricritical Points, in Phase 
Transitions and Critical Phenomena, Vol. 9, edited by C. Domb and J. L. Lebowitz, 
Academic Press, London, 1984.

\bibitem{RdaSilvaPRE2002} R. da Silva, N. A. Alves, e J. R. Drugowich de Fel%
\'{\i}cio, Universality and scaling study of the critical behavior of the
two-dimensional Blume--Capel model in short-time dynamics, Phys. Rev. E 
\textbf{66}, 026130 (2002)

\bibitem{RdaSilvaBJP2022} R. da Silva, Exploring the similarities between
mean-field and short-range relaxation dynamics of spin models, Braz. J.
Phys. \textbf{52}, (2022).

\bibitem{RdaSilvaPRE2022} R. da Silva, Numerical evidence of
Janssen-Oerding's prediction in a three-dimensional spin model far from
equilibrium, Phys. Rev. E \textbf{105}, 034114-1--034114-6 (2022)

\bibitem{RdasilvaIJMPC2023} R. da Silva, Random matrices theory elucidates
the nonequilibrium critical phenomena, Int. J. Mod. Phys. C \textbf{34},
2350061-1--2350061-12 (2023)

\bibitem{RdasilvaIJMPC2024} R. da Silva, E. Venites F., S. D. Prado, J. R.
Drugowich de Fel\'{\i}cio, Efficient Computational method using random
matrices describing critical thermodynamics, Int. J. Mod. Phys. C \textbf{37}%
, 2450163-1--2450163-24 (2024)

\bibitem{Eliseu2024} E .Venites F., R. da Silva, J. R. Drugowich de Fel\'{\i}%
cio, A Spectral Investigation of Criticality and Crossover Effects in Two
and Three Dimensions: Short Timescales with Small Systems in Minute Random
Matrices, Entropy \textbf{26(}5\textbf{)}, 395 (2024)

\bibitem{Lacasa2008} L. Lacasa, , B. Luque, F. Ballesteros, J. Luque, J. C.
Nu\~{n}o, From time series to complex networks: the visibility graph, PNAS 
\textbf{105}(13), 4972--4975(2008)

\bibitem{Zhao2017} L. Zhao, W. Li, C. Yang, J. Han, Z. Su, Y. Zou, A new
method for characterizing time series based on visibility graphs, PLoS ONE 
\textbf{12}(1), e0170467 (2017)

\bibitem{Gomez-Hernandez} D. G\'{o}mez-Hern\'{a}ndez , D. Garc\'{\i}a-Gudi%
\~{n}o, E. Landa , I. O. Morales , A. Frank, Characterization of time series
using visibility graphs: A comparative study, PLOS ONE \textbf{14}(9):
e0221674 (2019).

\bibitem{Ferreira} J. T. Moraes, S. C. Ferreira, Visibility graphs for
absorbing-state phase transitions, Phys. Rev. E \textbf{108}, 044309 (2023)

\bibitem{SilvaArxiv} R. da Silva, H. A. Fernandes, P. G. Freitas, S. Gon\c{c}%
alves, E. V. Stock, A. Alves, The number of spanning trees as an indicator
of critical phenomena: When Kirchhoff meets Ising, Phys. Rev. E (accepted
for publication) https://doi.org/10.1103/kzxs-9pv8 (2026)

\bibitem{Bollobas1998} B. Bollob\'{a}s, Modern Graph Theory, Springer, 1998.

\bibitem{Diestel2017} R. Diestel, Graph Theory, 5th ed., Springer, 2017

\bibitem{Stanley1999} R. P. Stanley, Enumerative Combinatorics, Vol. 2,
Cambridge University Press, 1999.

\bibitem{Kirchhoff1847}  G. Kirchhoff, On the solution of the equations to
which one is led in investigating the linear distribution of galvanic
currents, IRE Trans. Circuit Theory, \textbf{5}(1), 4--7 (1958), translated
by J. B. O'Toole from the original in German: \"{U}ber die Aufl\"{o}sung der
Gleichungen, auf welche man bei der Untersuchung der linearen Verteilung
galvanischer Str\"{o}me gef\"{u}hrt wird, Annalen der Physik und Chemie, 
\textbf{72}, 497--508 (1847).

\bibitem{Harary1969} Frank Harary, Graph Theory, Addison-Wesley, Reading, 
Massachusetts, 1969.

\bibitem{Butera} P. Butera, M. Pernici, The Blume--Capel model for spins S =
1 and 3/2 in dimensions d = 2 and 3 Physica A \textbf{507,} 22--66 (2018)

\bibitem{DeBeale} P.D. Beale, Finite-size scaling study of the
two-dimensional Blume--Capel model, Phys. Rev. B \textbf{33}, 1717--1720
(1986)

\bibitem{Oganesyan} V. Oganesyan, D. Huse, Localization of interacting
fermions at high temperature, Phys. Rev. B \textbf{75}, 155111-1--155111-5
(2007)

\bibitem{Atas} Y.Y. Atas, E. Bogomolny, O. Giraud, G. Roux, Phys. Rev. Lett. 
\textbf{110}, 084101-1--084101-5 (2013)
\end{thebibliography}
\end{document}